\pdfoutput=1
\RequirePackage{lineno}
\newif\ifpreprint%
\preprintfalse%
\ifpreprint%
	\documentclass[preprint,english,aps,prb,floatfix,amssymb]{revtex4-1}
\else%
	\documentclass[twocolumn,english,aps,prb,floatfix,amssymb]{revtex4-1}
\fi%
\usepackage{amsmath}
\usepackage{dsfont}
\usepackage{color}
\usepackage{amssymb}
\usepackage{graphicx}
\usepackage{braket}
\usepackage{units}
\usepackage{lipsum}
\ifpreprint%
\else%
\fi%

\usepackage{hyperref}
\hypersetup{
  colorlinks = true,
  hypertexnames=true
}
\newcommand{\ssm}{\scriptscriptstyle\rm}

\renewcommand{\theta}{\vartheta}
\renewcommand{\phi}{\varphi}

\begin{document}
\ifpreprint%
	\linenumbers%
\fi%

\title{Anomalous Fermi arcs in a periodically driven Weyl system}

\author{
Valerio Peri}
\affiliation{
 Institute for Theoretical Physics, ETH Z\"urich, 8093 Zurich, Switzerland
}

\author{
Sebastian D. Huber}
\affiliation{
  Institute for Theoretical Physics, ETH Z\"urich, 8093 Zurich, Switzerland
}

\begin{abstract}
Three dimensional Weyl semimetals exhibit open Fermi arcs on their sample surfaces connecting the projection of bulk Weyl points of opposite chirality. The canonical interpretation of these surfaces states is in terms of chiral edge modes of a layer quantum Hall effect: The two-dimensional momentum-space planes perpendicular to the momentum connecting the two Weyl points are characterized by a non-zero Chern number. It might be interesting to note, that in analogy to the known two-dimensional Floquet anomalous chiral edge states, one can realize open Fermi arcs in the absence of Chern numbers in periodically driven system. Here, we present a way to construct such anomalous Fermi arcs in a concrete model.
\end{abstract}

\date{\today}

\maketitle
Topological band theory has provided us with a helpful classification of insulators and semimetals.\cite{Kitaev:2009,Schnyder:2008,Bradlyn:2017,Chiu:2014,Shiozaki:2014} Moreover, much of our understanding of intriguing electronic phenomena has been influenced by simple concepts such as the bulk-edge correspondence in the quantum Hall effect: A non-zero bulk Chern number of a two-dimensional system ensures the presence of one-dimensional chiral edge states. \cite{Thouless:1982,Laughlin:1981,Avron:1985}

\begin{figure}[tbh]
\includegraphics{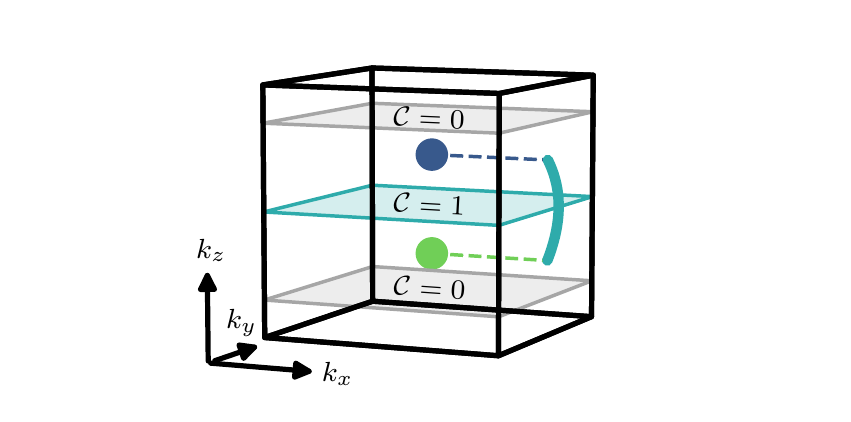}
\caption{\label{fig:fig1}Schematic representation of a Weyl system. Weyl points of positive (green) and negative (blue) chirality are present in the three-dimensional bulk. On the surface, open equienergy lines (light blue) connect the projection of the Weyl points. These arcs are a manifestation of non-trivial layer Chern number $\mathcal{C}$ per momentum space slice. In a periodically driven system, such a situation can be enriched by the anomalous Fermi arcs described in this Communication.}
\end{figure}

Weyl semimetals are a prime example of band-strucutres that can be understood by reducing them to a set of quantum Hall systems.\cite{Armitage:2018,Burkov:2016,Yoshimura:2016,Burkov:2011} Traversing a conical touching point in momentum space changes the Chern number of the two-dimensional slices perpendicular to the direction in which the Weyl point is crossed. This is most easily seen in the example 
\begin{equation}
\label{eq:simpleModel}
  H=v_xk_x\sigma_x+v_yk_y\sigma_y+v_z k_z\sigma_z.
\end{equation} 
Here $\sigma_\alpha$ denote the standard Pauli matrices and $v_\alpha$, the group velocities. We know from standard manipulations\cite{Haldane:1988} that the Chern numbers for the $k_x$--$k_y$ layers are given by $\frac{1}{2}{\rm sign}(v_zk_z)$. In other words, the layer Chern number changes by $\text{sign}(v_z)$ when we ramp $k_z$ from negative to positive values. Consequently, the number of chiral edge channels per momentum space layer is changing. This abrupt change in the number of surface states is manifested by the open Fermi arcs. A schematic representation of this situation is presented in Fig.~\ref{fig:fig1}. In the more generic case
\begin{equation}
  H=\sum_{\alpha,\beta}v_{\alpha\beta} k_\alpha\sigma_\beta,
\end{equation}
it is the chirality $s=\text{sign}[\text{det}(v_{\alpha\beta})]=\pm 1$ of the Weyl point that determines the sign of the change in the layer Chern number.

Recent advances in our understanding of periodically driven systems\cite{Oka:2008,Kitagawa:2010,Lindner:2011,Roy:2017,Yao:2017a} have revealed that protected chiral modes on the edges of two-dimensional systems do not have to be linked to any of the known topological quantum numbers.\cite{Rudner:2013} In fact, there are configurations of bands and in-gap states that can only be realized in Floquet systems.\cite{Rudner:2013,Maczewsky:2017,Mukherjee:2017,Peng:2016} In this Communication we show how one can obtain open Fermi arcs using these anomalous edge modes by stacking Floquet topological insulators into a three dimensional Weyl system. In other words, we propose a simple way to obtain open Fermi arcs that are not described by layer Chern numbers.

In periodically driven systems, the Hamiltonian $H(t)=H(t+T)$ defines the period $T$. The Floquet states  $\ket{\psi(t)}$ are eigenstates of $H(t)$ and satisfy the relation $\ket{\psi(T)}=e^{-i\epsilon T}\ket{\psi(0)}$, where the phase $\epsilon$ is commonly called quasi-energy. Importantly, as it controls the phase evolution over one period $T$, it is periodic $\epsilon=\epsilon+n\omega$, with $n \in \mathbb{Z}$ and $\omega=2\pi/T$. This property parallels the periodicity of the crystal momentum in lattice systems and it is defined in the quasi-energy Brillouin zone $\epsilon \in [-\omega/2, \omega/2]$. A set of bands $\{\epsilon_l({\bf k})\}$  describes the Floquet spectrum that replaces the energy spectrum of static systems. 

The $\omega$-periodicity of the spectrum is at the core of anomalous behaviours predicted in Floquet systems that have no static counterparts.\cite{Franca:2018,Huang:2018,Ladovrechis:2018,Leykam:2016,Higashikawa:2018,Zhou:2018a} For example, it was realized in Ref.~\onlinecite{Rudner:2013} that in two-dimensional systems anomalous surface states can appear even when all Floquet bands have a zero Chern number.

The Floquet formalism has recently been applied to Weyl semimetals. For example, the possibility of observing a single Weyl point without a partner of opposite chirality $s$ as been proposed.\cite{Higashikawa:2018, Sun:2018} A situation forbidden by the Nielsen-Ninomiya theorem in static systems.\cite{Nielsen:1983} Weyl points have been predicted to emerge from topolgoical insulators and Dirac semimetals in the presence of an external periodic drive.\cite{Wang:2014,Zou:2016,Zhang:2016a,Hubener:2017,Bucciantini:2017} Here we promote the idea that open Fermi arcs can be obtained without layer Chern numbers. 

We start with the time dependent Hamiltonian
\begin{equation}
  H(t)=\sum_{\alpha,\alpha',{\bf k}}c^{\dagger}_{\alpha{\bf k}}H_{\alpha\alpha'}({\bf k},t)c_{\alpha'{\bf k}}\,,
\end{equation}
where $c^\dagger_{\alpha k}$ ($c_{\alpha k}$) creates (annhilates) an electron in the orbital $\alpha$ with momentum ${\bf k}$. If we expand the Floquet states of the band $l$ in a basis $\ket{\psi_l({\bf k},t)}=\sum_\alpha \Phi_{l\alpha}({\bf k},t)c^\dagger_{\alpha{\bf k}}\ket{0}$, the amplitudes $\Phi_{l\alpha}({\bf k},t)$ obey the Schr\"odinger equation 
\begin{equation}
  i\partial_t\Phi_{l\alpha}({\bf k},t)=\sum_{\alpha'}H_{\alpha\alpha'}({\bf k}\,,t)\Phi_{l\alpha'}({\bf k},t)\,.
  \end{equation}
Using the Floquet theorem we can rewrite the amplitudes as
\begin{equation}
  \Phi_{l\alpha}({\bf k},t)=e^{-i\epsilon_n({\bf k}) t}\sum_{m=-\infty}^{+\infty}e^{im\omega t}\varphi^{(m)}_{l\alpha}({\bf k}).
\end{equation}
Thus, we are left with a time independent eigenvalue problem for the coefficients $\varphi^{(m)}_{l\alpha}({\bf k})$
\begin{equation} \sum_{\alpha',m'}\mathcal{H}_{\alpha\alpha'}^{mm'}({\bf k},\omega)\varphi^{(m')}_{l\alpha'}({\bf k})=\epsilon_l\varphi^{(m)}_{l\alpha}({\bf k}).
  \end{equation}
Here, $\mathcal{H}_{\alpha\alpha'}^{mm'}$ are the Fourier coefficients 
\begin{equation} \mathcal{H}_{\alpha\alpha'}^{mm'}({\bf k},\omega) = \frac{1}{T}\int_0^T dt\,
        e^{2\pi it(m-m')/T}H({\bf k},t).
\end{equation}
In principle, the spectrum extends from $-\infty$ to $+\infty$. However, we can restrict our attention to quasi-energies $\epsilon_l({\bf k})\in [-\omega/2,\omega/2]$ owing to the periodicity of the Floquet spectrum.

\begin{figure}[tbh]
\includegraphics{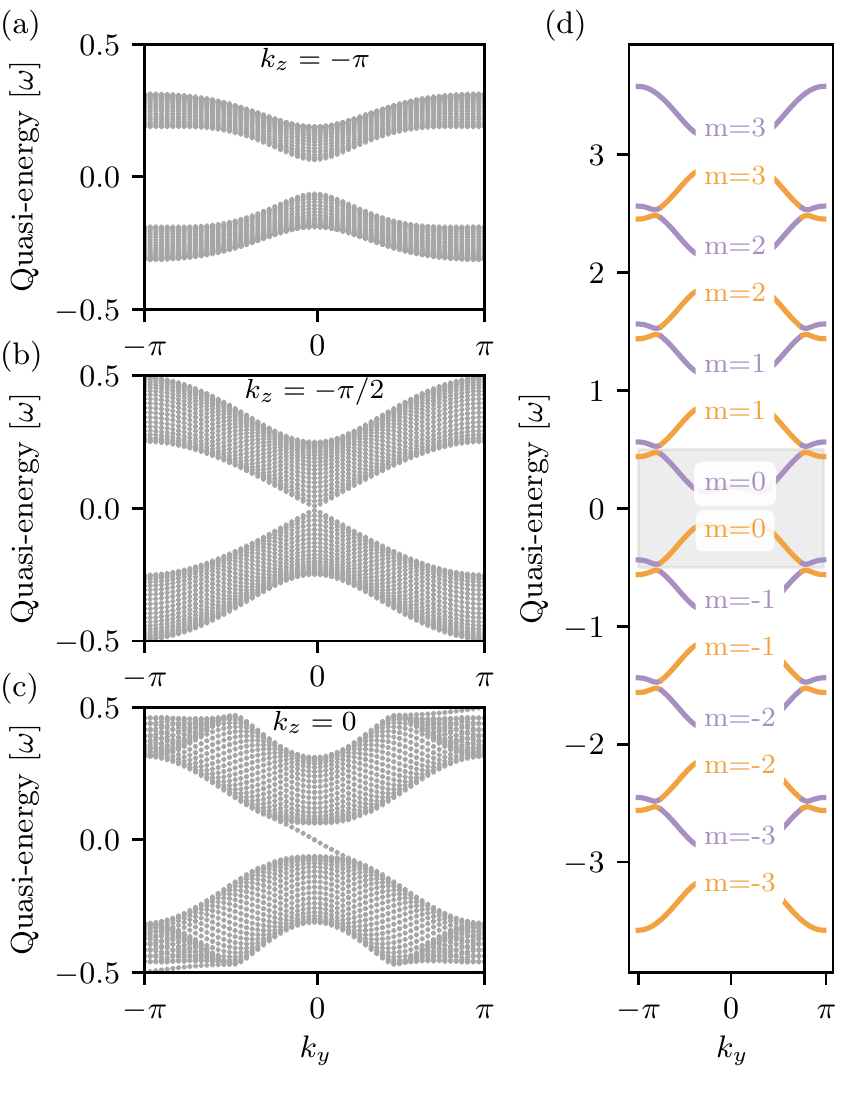}
\caption{\label{fig:fig2}(a) Floquet spectrum of the model of Eq.~\eqref{eq:model} for $k_z=-\pi$. (b) Same as (a) but for $k_z=-\pi/2$. Both the gaps at $0$ and at $\pi/T$ close. (c) For $k_z=0$ Fermi arcs localized at the surface appear in the middle of the gaps. Only the arcs on one surface have been plotted. The model parameters are $b_0=1$, $b=0.5$, $\mu=1$, $\Delta=2$ and $\omega=16b_0$ in a finite size system $L=20$ with open boundaries.(d) Truncated Floquet spectrum at $M=4$ for the same model. The parameters are the same but the system is assumed to be periodic. The spectrum is computed for $k_z=0$ along the line $k_x=k_y$.}
\end{figure}

We assume the following form of the Hamiltonian
\begin{equation}
  H(t)=H_0+\Omega e^{i\omega t}+\Omega^\dagger e^{-i\omega t},
\end{equation}
where $H_0$ is the time-independent part. In this case of a harmonic driver, $\mathcal{H}_{\alpha\alpha'}^{mm'}$ assumes a tri-diagonal form
\begin{equation}
  \label{eq:eF}
  \mathcal{H}({\bf k},\omega)=\begin{pmatrix}
  \ddots & & & &\\
  & H_0+\omega\mathds{1} &  \Omega & 0 &\\
  & \Omega & H_0 & \Omega &\\
  & 0 & \Omega & H_0-\omega\mathds{1}& \\
  & & & & \ddots \\
  \end{pmatrix}\;.
  \end{equation}
The infinite ladder of detuned $H_0$'s is equivalent to the Wannier-Stark problem:\cite{Wannier:1960} The Floquet eigenstates are localized around one copy with fixed $m_0$ and decay exponentially with the distance $|m-m_0|$. This allows to truncate the infinite ladder and to consider only $m$ and $m'$ such that $-M\leq m\,,m'\leq M$, where $M$ depends on the driving strength $\Omega$. In Fig.~\ref{fig:fig2}(d), the truncated spectrum with $M=4$ for the model of Eq.~\eqref{eq:model} is shown.

In a driven two-band system, there will be generically two band gaps: One around $\epsilon=0$ arising from the original static Hamiltonian $H_0$ and one around $\epsilon=\omega/2$ induced by the drive $\Omega$. A similar situation has been analyzed in Ref.~\onlinecite{Rudner:2013} for anomalous edge states in two-dimensional systems. Here, we consider driven two-dimensional systems stacked along a third dimension, such that the gaps close and reopen as a function of $k_z$. The key idea of this Communication is that if the gap closings at $\epsilon=0$ and $\epsilon=\omega/2$ happen for the same value of $k_z$, one might induce a quantum of Berry flux of opposite sign on the two opposite band edges, and hence leave the Chern number untouched. However, the number of edge states will be changed by one for each of the two gaps. We now present a concrete model that illustrates this idea.

The static Hamiltonian is given by 
\begin{equation}
        H_0=\sum_{\alpha=x,y,z} d_\alpha({\bf k}) \sigma_\alpha 
        \label{eq:model}
\end{equation}
with
\begin{align}
\label{eq:3dD}
d_x({\bf k})&=a\sin{k_x}\,, \quad \quad d_y({\bf k})=a\sin{k_y},\\
d_z({\bf k})&=\mu\cos{k_z}-2(b_0+b\cos{k_z})(2-\cos{k_x}-\cos{k_y}).
\nonumber
\end{align}
We assume $a,\mu,b_0,b$ to be positive. For $\mu_{\ssm eff}=\mu\cos(k_z)$ and $b_{\ssm eff}=b_0+b\cos(k_z)$, with $b<b_0$, the above Hamiltonian describes the usual lattice Dirac model with a gap-closing transition at $k_x=k_y=0$ for $\mu_{\ssm eff}=0$. In other words, at ${\bf k}=(0,0, \pm\pi/2)$, $H_0$ has a Weyl point at $\epsilon=0$. By expanding ${\bf d}({\bf k})$ around these points we find that 
\begin{equation}
        v_{\alpha\beta}=\text{diag}(a,a,\mp \mu)
\end{equation}
for $k_z=\pm \pi/2$. Therefore the chiralities of the Weyl points are given by $s=\mp 1$ for ${\bf k}=(0,0,\pm \pi/2)$.

In order to avoid a non-zero Chern number induced by the gap closings at $\epsilon=0$, we drive the system in a way that we remove the Berry curvature induced by $H_0$ through the gap closing at $\epsilon=\omega/2$. We assume 
\begin{align}
        \label{eq:small}
         b_0&> b-\mu/4\quad \mbox{and} \quad \mu<8b, \\
         \omega&=16b_0,\\
         \Omega&= \Delta \sigma_z \quad \mbox{with} \quad |\Delta |< b_0.
\end{align}
The first line ensures that the upper band edge of $H_0$ is located at $k_x=k_y=\pi$ with a bandwidth of $16b_0$ at $k_z=\pm \pi/2$. By choosing $\omega=16b_0$, we have a quadratic band touching at ${\bf k}=(\pi,\pi,\pm \pi/2)$ at the quasi-energy zone boundary. Second-order degenerate perturbation theory in $\Delta/b_0$ reveals that the low-energy theory at the quasi-energy zone boundary develops from a quadratic band touching into a Weyl point with 
\begin{equation}
       v_{\alpha\beta}=\text{diag}\left[
        -\frac{\Delta a}{8b_0},
        -\frac{\Delta a}{8b_0},
        \mp(-8b+\mu)
       \right].
\end{equation}
Condition (\ref{eq:small}) ensures that the chiralities $s=\pm 1$ at ${\bf k}=(\pi,\pi,\pm \pi/2)$ and hence they exactly cancel the effect on the Chern number of the Weyl points at $\epsilon=0$.

The setup of our model (\ref{eq:3dD}) is fine-tuned to have the two Weyl points at $\epsilon=0$ and $\epsilon=\omega/2$ at the same $k_z$ value. If, e.g., the driving frequency is slightly detuned from $16b_0$, the Weyl points at $\epsilon=0$ and $\epsilon=\omega/2$ do not appear for the same $k_z$. While this induces a small region in $k_z$, where the layer Chern numbers are non-zero, the strict link between layer Chern numbers and open Femi arcs is still broken. 

In Fig.~\ref{fig:fig2} we present the evolution of the Floquet spectrum as a function of $k_z$. Fig.~\ref{fig:fig2}(a) shows the bands at $k_z=-\pi$ where there is a gap both at $\epsilon=0$ and $\epsilon=\pi/T$. At $k_z=-\pi/2$, the gaps simultaneously close, cf. Fig.~\ref{fig:fig2}(b). Fermi arcs appear for $k_z=0$, as shown in Fig.~\ref{fig:fig2}(c). These are anomalous Fermi arcs that cannot be understood in terms of layer Chern number.

Fermi arcs have been shown to be sensitive to disorder.\cite{Wilson:2018,Slager:2017} This is in striking  contrast to edge states of Chern insulator that are protected by the bulk gap. Anomalous Floquet Anderson insulators, on the other hand, are known to evolve into a different phase than their un-driven disordered counterparts.\cite{Titum:2016,Kundu:2017,Nathan:2017,Liu:2018} It is an interesting question for further investigation if the anomalous Fermi arcs presented here react differently to disorder than their standard Weyl counterparts.

\vspace{.5cm}

We acknowledge insightful discussions with Erez Berg and Titus Neupert. We are grateful for the financial support from the Swiss National Science Foundation, the NCCR QSIT, and the ERC project TopMechMat.

\bibliography{ref}
\bibliographystyle{phd-url}
\clearpage

\end{document}